\documentclass[conference]{IEEEtran}

\usepackage[cmex10]{amsmath}

\usepackage{graphicx} %graphics
\usepackage{graphics}
\usepackage[font=footnotesize]{caption}
\usepackage{subcaption}
\usepackage{epstopdf}
\usepackage{enumitem}
\usepackage{color}
\usepackage{float}
\usepackage{amsmath} %Equations
\usepackage{amssymb}
\usepackage{amsthm}
\usepackage{algorithm}
\usepackage{algorithmic}
\usepackage{multirow}
\usepackage{multicol}
\usepackage{cite}

\theoremstyle{plain}

\newtheorem{lemma}{Lemma}

\IEEEoverridecommandlockouts

\newcommand{\vect}[1]{\boldsymbol{\mathbf{#1}}}
\newcommand{\norm}[1]{\left\|#1\right\|}
\newcommand{\dimn}[1]{\in \mathbb{C}^{#1}}
\renewcommand{\gamma}{\mu}

\DeclareMathOperator{\diag}{diag}
\DeclareMathOperator{\tr}{tr}
\def\Htran{\mbox{\tiny $\mathrm{H}$}}
\def\Ttran{\mbox{\tiny $\mathrm{T}$}}
\def\CN{\mathcal{N}_{\mathbb{C}}} %Complex Gaussian

\begin{document}

\title{Centralized and Distributed Power Allocation for Max-Min Fairness in Cell-Free Massive MIMO\vspace{-0.3cm}}

\author{Sucharita Chakraborty\IEEEauthorrefmark{1}, Emil Bj{\"o}rnson\IEEEauthorrefmark{1}, and Luca Sanguinetti\IEEEauthorrefmark{2}\\
	\IEEEauthorrefmark{1}Department of Electrical Engineering (ISY), Link{\"o}ping University, \\ Link{\"o}ping, Sweden (\{sucharita.chakraborty, emil.bjornson\}@liu.se), \\
	\IEEEauthorrefmark{2}Dipartimento dell$\hat{\textrm{a}}$Informazione, University of Pisa, 56122 Pisa, Italy (luca.sanguinetti@unipi.it)\thanks{This work was partially supported by the Wallenberg AI, Autonomous Systems and Software Program (WASP) funded by the Knut and Alice Wallenberg Foundation.}}

\maketitle

\begin{abstract}
Cell-free Massive MIMO systems consist of a large number of geographically distributed access points (APs) that serve users by coherent joint transmission. Downlink power allocation is important in these systems, to determine which APs should transmit to which users and with what power. If the system is implemented correctly, it can deliver a more uniform user performance than conventional cellular networks. To this end, previous works have shown how to perform system-wide max-min fairness power allocation when using maximum ratio precoding. In this paper, we first generalize this method to arbitrary precoding, and then train a neural network to perform approximately the same power allocation but with reduced computational complexity. Finally, we train one neural network per AP to mimic system-wide max-min fairness power allocation, but using only local information. By learning the structure of the local propagation environment, this method outperforms the state-of-the-art distributed power allocation method from the Cell-free Massive MIMO literature.~
\end{abstract}


\vspace{-0.2cm}

\IEEEpeerreviewmaketitle

\section{Introduction}
Coordinated distributed wireless systems liberate the conventional co-located multiple-input multiple-output (MIMO) from its shackles of inherent form-factor constraint \cite{Interdonato2018}. If an arbitrarily large number of collaborative access points (APs) jointly serve the users in a wide area, it constitutes a dense large network without any cell boundaries. This type of systems is gaining popularity with the name of \emph{Cell-free massive MIMO (mMIMO) systems} \cite{ngo2017cell,bjornson2019making, nayebi2017precoding}. It reaps many of the advantages of two cornerstone technologies: mMIMO (e.g., favorable propagation) and Network MIMO (e.g., more uniform user performance) by exploiting coherent signal co-processing among multiple distributed APs. 

The main challenge in bringing such a network to reality is scalability, in terms of computational complexity for signal processing and resource allocation, fronthaul requirements, etc. General guidelines for Network MIMO were provided in \cite{bjornson2013optimal} and later particularized for Cell-free mMIMO in \cite{bjornson2019scalable,interdonato2019scalability}.
Existing algorithms in Network MIMO, or coordinated multipoint (CoMP), are mainly limited by three factors \cite{bjornson2019scalable}:
\begin{enumerate}[leftmargin=*]
\item Dependency on the availability of full channel state information (CSI) in the network, or at least partial CSI that is shared between neighboring APs. In future ultra dense networks, where the number of APs grows large, significant communication overhead will be incurred in acquiring and disseminating CSI to all of the cooperative APs or to an edge cloud computer. 

\item An enormous amount of information regarding the payload data (e.g., coding/modulation scheme, decoding errors) must be monitored to guarantee user satisfaction. Immense buffer sizes and fronthaul signaling capacity are required to store and share such global information over the network. 

\item The time delay and complexity incurred during channel estimation, precoding/combining, and fronthaul signaling increase at least linearly with the number of APs.
\end{enumerate}
Most of these issues were overlooked in the early works on Cell-free mMIMO \cite{ngo2017cell, nayebi2017precoding}. However, the recent works \cite{bjornson2019scalable,interdonato2019scalability} have indicated that a scalable implementation might be realizable in practice. This paper focuses on the scalable implementation of downlink power allocation algorithms.

Network-wide downlink power allocation algorithms for Cell-free mMIMO systems were developed in \cite{nayebi2017precoding,nayebi2016performance} for the purpose of achieving max-min fairness; that is, all user equipments (UEs) get the same spectral efficiency (SE) and that common value is maximized. The results apply to maximum ratio (MR) precoding with long-term power constraints, which are undesirable assumptions since the use of regularized zero-forcing (RZF) precoding at every AP gives higher SE \cite{bjornson2019scalable} and real systems are subject to short-term power constraints \cite{Interdonato2016a}.

Even in the case when these algorithms are applicable, the deployment feasibility is limited since global CSI must be available at a central processing unit (CPU) and the computational complexity grows polynomially with the number of APs and UEs. The first of these issues can be addressed by utilizing the dynamic cooperation cluster (DCC) concept \cite{bjornson2013optimal,bjornson2019scalable}, in which each user is assumed to be served by a user-centrically selected subset of the APs with the best channel conditions. In this paper, we first develop an optimal power allocation algorithm for max-min fairness in DCC-based Cell-free mMIMO systems, but the complexity is unscalable. We then utilize the \emph{learn to optimize} framework \cite{sun2018learning,sanguinetti2018deep} for offline training of deep neural networks (DNNs) that perform approximated max-min fairness power allocation. 

More precisely, we train one DNN to perform centralized power allocation with reduced computational complexity. We also train one DNN per AP to perform distributed power allocation using only locally available information as input, but training it using the globally optimal max-min fairness solution. Hence, different from the heuristic power allocation method that was recently proposed in \cite{interdonato2019scalability}, each AP is utilizing a unique algorithm that can take the actual network structure and propagation environment into account.

\textbf{Notations:} $x$, $\vect{x}$, $\vect{X}$ denote a scalar, vector, and matrix, respectively. $(\cdot)^{\Ttran}$, $(\cdot)^{\Htran}$, $|\cdot|$, $\| \cdot \|$, $\vect{I}_{X}$ stand for transpose, Hermitian transpose, absolute value, the $L_{2}$ vector norm, and $X\times X$ identity matrix, respectively. The multivariate circular symmetric complex Gaussian distribution with correlation matrix $\vect{R}$ is denoted $\CN(\vect{0},\vect{R})$.

\section{System Model}

We consider a Cell-free mMIMO system with $K$ single-antenna UEs and $L$ APs, each equipped with $N$ antennas. We assume a block fading channel model where the channels are static within time-frequency coherence blocks with $\tau_c$ channel uses, and independent random channel realizations appear in each block.
The channel between the $k$th UE and the $l$th AP is denoted as $\vect{h}_{kl}\in \mathbb{C}^{N \times 1}$ and is modeled by correlated Rayleigh fading $\vect{h}_{kl}\sim \CN (\vect{0}, \vect{R}_{kl})$,
where $\vect{R}_{kl}\in \mathbb{C}^{N\times N}$ is the spatial correlation matrix. The normalized trace $\beta_{kl} = {1}/{N} \tr ( \vect{R}_{kl})$
accounts for the average channel gain from an antenna at AP~$l$ to UE~$k$.

The APs are connected  to a CPU via fronthaul connections, which are used to convey uplink and downlink data between the APs and the CPU. The connections are assumed to be error-free but the capacity is limited, thus each UE can only be served by a subset of the APs. No instantaneous CSI is conveyed over the fronthaul.

\subsection{Channel Estimation}
We consider a time division duplex (TDD) protocol having a pilot transmission phase for channel estimation and a data transmission phase. Following the standard TDD protocol \cite{Interdonato2018}, the coherence block is divided in three parts: $\tau_{p}$ for uplink pilot transmission, $\tau_{u}$ for uplink data transmission, and $\tau_{d}$ for downlink data. It thus follows that $\tau_{c} = \tau_{p} + \tau_{u} + \tau_{d}$. 

In the uplink pilot phase, a set of $\tau_{p}$ mutually orthogonal $\tau_{p}$-length pilots are utilized. Each UE is assigned to one of these pilots.
Let us denote the subset of UEs that are assigned to pilot $t$ as $\mathcal{P}_{t}\subset \{1,\ldots,K\}$. The received signal $\vect{y}_{tl}^{p}\dimn{N\times 1}$ at AP $l$ when the UEs in $\mathcal{P}_{t}$ transmit is defined as
\begin{equation}
	\vect{y}_{tl}^{\textrm{p}}=\sum_{i\in \mathcal{P}_{t}}\sqrt{\tau_{p}p_{i}}\vect{h}_{il} + \vect{n}_{tl}
\end{equation}
where $p_{i}$ is the transmit power from the $i$th UE and $\vect{n}_{tl} \sim \CN (\vect{0},\sigma^2 \vect{I}_N)$ is the additive white Gaussian noise (AWGN) with variance $\sigma^2$. By utilizing the standard minimum mean square error (MMSE) estimator at the $l$th AP, the estimate of the channel $\vect{h}_{kl}$ from UE $k \in \mathcal{P}_t$ is \cite{bjornson2019making} 
\begin{align}
	\hat{\vect{h}}_{kl}&=\sqrt{\tau_{p}p_{k}}\vect{R}_{kl}\left(\sum_{i\in \mathcal{P}_{t}}\tau_{p}p_{i}\vect{R}_{il} + \sigma^2 \vect{I}_{N}\right)^{-1}\vect{y}_{tl}^{\textrm{p}} \nonumber \\
	&\sim  \CN \left( \vect{0}, \tau_{p}p_{k}\vect{R}_{kl}\vect{\Phi}_{kl}^{-1}\vect{R}_{kl} \right)
\end{align}
where $\vect{\Phi}_{kl}=\mathbb{E}\{\vect{y}_{tl}^{p}\left(\vect{y}_{tl}^{p}\right)^{\Htran}\} = \sum_{i\in \mathcal{P}_{t}}\tau_{p}p_{i}\vect{R}_{il} + \sigma^2 \vect{I}_{N}$ denotes the correlation matrix of the pilot signal.

\subsection{DCC Framework Based Cell-free mMIMO}

We assume that each AP serves a subset of the UEs and we use the DCC framework \cite{bjornson2013optimal,bjornson2019scalable}. We let $\mathcal{D}_{l} \subset \{1, \ldots,K\}$ denote the UEs served by the $l$th AP. 
In accordance to \cite{bjornson2013optimal}, we then define the matrices $\vect{D}_{kl} \in \mathbb{C}^{N \times N}$, for $l=1,\ldots,L$ and $k=1,\ldots,K$, as
\begin{equation}
	\vect{D}_{kl}=\begin{cases}
	\vect{I}_{N} & \text{for } k\in \mathcal{D}_l, \\
	\vect{0}_{N} & \text{for } k \notin \mathcal{D}_l.
	\end{cases}
\end{equation}
Notice that this matrix is $\vect{I}_{N} $ if the $k$th UE is served by the $l$th AP and $\vect{0}_{N}$ otherwise. 
The received downlink signal at the $k$th UE reads as
\begin{align} \label{eq:downlink-set-notation}
y_{k}^{\mathrm{dl}} &=\sum_{l=1}^{L} \vect{h}_{kl}^{\Htran}\sum_{i \in \mathcal{D}_l} \sqrt{\rho_{il}} \vect{w}_{il}s_{i}+n_{k} \\ 
&= \sum_{l=1}^{L}\vect{h}_{kl}^{\Htran}\sum_{i=1}^{K}\sqrt{\rho_{il}} \vect{D}_{il}\vect{w}_{il}s_{i}+n_{k} \label{eq:downlink-DCC-notation}
\end{align}
where $\rho_{il}$ is the downlink power allocated to UE $i$ by AP $l$ and $\vect{w}_{il}\in \mathbb{C}^{N\times 1}$ is the corresponding normalized precoding vector with $\|\vect{w}_{il}\|^2=1$. Moreover, $s_{i}$ denotes the signal transmitted to UE $i$ and $n_{k}\sim  \CN (0, \sigma^{2})$ is the receiver noise.
 
The benefit of using the matrix notation in \eqref{eq:downlink-DCC-notation}, instead of the set notation in \eqref{eq:downlink-set-notation}, is that sizes of all matrices and vectors become independent of which APs serve which UEs. This will be convenient in Section~\ref{sec:downlink-power-allocation}. The original Cell-free mMIMO model in \cite{ngo2017cell, nayebi2017precoding} is obtained by setting $\mathcal{D}_l = \{ 1, \ldots, K \}$ and thus $\vect{D}_{kl} = \vect{I}_N$ for all $l$ and $k$. Practical methods to select the sets $\mathcal{D}_1, \ldots, \mathcal{D}_L$ are found in \cite{bjornson2013optimal,bjornson2019scalable}.

\subsection{Downlink SE}

The downlink SE of Cell-free mMIMO was characterized for MR precoding with long-term power constraints in \cite{ngo2017cell, nayebi2017precoding}. This enabled the development of max-min power allocation optimization. The SE with arbitrary precoding schemes was considered in \cite{bjornson2019scalable}, but the expression was not amendable for power optimization. The following lemma provides a new simplified expression for the general case, which will enable the power optimization in Section~\ref{sec:downlink-power-allocation}.

\begin{lemma} \label{lemma:SE}
An achievable downlink SE for UE $k$ is
\begin{equation}
	\mathrm{SE}_{k}^{\mathrm{dl}} = \frac{\tau_{d}}{\tau_{c}}\log_{2}\left(1+\mathrm{SINR}_{k}^{\mathrm{dl}}\right)
\end{equation}
where
	\begin{align}\notag
	&\mathrm{SINR}_{k}^{\mathrm{dl}} = \\
	 &\frac{\left\lvert\sum\limits_{l=1}^{L}\sqrt{\rho_{kl}}\mathbb{E}\{\vect{h}_{kl}^{\Htran}\vect{D}_{kl}\vect{w}_{kl}\}\right\rvert^{2}}{ \sum\limits_{l=1}^{L}\sum\limits_{i=1}^{K}\rho_{il}\mathbb{E}\left\{\left\lvert\vect{h}_{kl}^{\Htran}\vect{D}_{il}\vect{w}_{il} \right\rvert^{2} \right\}\!-\!\sum\limits_{l=1}^{L}\rho_{kl}\left\lvert\mathbb{E}\{\vect{h}_{kl}^{\Htran}\vect{D}_{kl}\vect{w}_{kl}\}\right\rvert^{2}\!+\!\sigma^{2}}\label{eq:SINR definition}
	\end{align}
	is the effective signal-to-interference-and-noise ratio (SINR).
\end{lemma}

The SE expression in Lemma~\ref{lemma:SE} can be utilized along with any precoding scheme and correlated Rayleigh fading model.\footnote{Actually, it holds for any arbitrary independent fading distribution.} 
We consider precoding vectors $\{\vect{w}_{il}\}$ satisfying short-term power constraints, which means that $\|\vect{w}_{il}\|^2=1$ must be satisfied in every coherence block and not on average (i.e., $\mathbb{E}\{\|\vect{w}_{il}\|^2\}=1$) as in \cite{ngo2017cell, nayebi2017precoding}. This is the conventional approach to precoding normalization \cite{bjornson2013optimal}. We notice that the relaxed long-term average power constraints are popular in Massive MIMO because they lead to closed-form SINR expressions. The relaxation is rather tight in Massive MIMO since the channel hardening makes $\|\vect{w}_{il}\|^2 \approx \mathbb{E}\{\|\vect{w}_{il}\|^2\}$. However, the same relaxation should not be used in Cell-free mMIMO since the channel between an AP with few antennas and a UE does not harden (although the joint channel from all APs might harden in some cases).

An arbitrary normalized precoding vector is defined as $\vect{w}_{kl}={\bar{\vect{w}}_{kl}}/{\norm{\bar{\vect{w}}_{kl}}}$
where $\bar{\vect{w}}_{kl}$ can be arbitrarily selected. In this paper, we consider MR and RZF precoding, which are defined as
\begin{equation}
	\bar{\vect{w}}_{kl}=\begin{cases}
	\hat{\vect{h}}_{kl}  & \text{for MR}, \\
	\left(\sum\limits_{i \in \mathcal{D}_{l}}p_{i}\hat{\vect{h}}_{il}\hat{\vect{h}}_{il}^{\Htran} + \sigma^{2}\vect{I}_{N}\right)^{-1}p_{k}\hat{\vect{h}}_{kl} & \text{for RZF}.\!\!\!
	\end{cases}
\end{equation}

\section{Downlink Max-Min Power Allocation}
\label{sec:downlink-power-allocation}

In this section, we generalize the max-min fairness algorithm from \cite{ngo2017cell, nayebi2017precoding} to general precoding schemes and correlated Rayleigh fading.
Hence, the goal is to find the optimal power allocation coefficients $\{\rho_{kl}: \forall k,l\}$ that maximize the lowest SE among all UEs in the network. Hence, we select the coefficients to give all UEs the same effective SINR and this value is to be maximized, under the constraint that each AP has the same maximum power $P^{\mathrm{dl}}_{\max}$. This means that the power constraint at AP $l$ is $\sum_{k=1}^{K} \rho_{kl} \leq P^{\mathrm{dl}}_{\max}$.

Before formulating the max-min fairness optimization problem, we first rewrite (\ref{eq:SINR definition})  as
\begin{equation}\label{eq:Redefined SINR}
\mathrm{SINR}_{k}^{\mathrm{dl}} = \frac{\left\lvert\sum\limits_{l=1}^{L}\gamma_{kl} a_{kl} \right\rvert^{2}}{ \sum\limits_{l=1}^{L}\sum\limits_{i=1}^{K}\gamma_{il}^{2} b_{kil}+\sigma^{2}}
\end{equation}
by introducing the new variables $\gamma_{kl} = \sqrt{\rho_{kl}}$ and 
\begin{align}
	a_{kl} &= \mathbb{E}\{\vect{h}_{kl}^{\Htran}\vect{D}_{kl}\vect{w}_{kl}\}\\
	b_{kil} &= \mathbb{E}\left\{\left\lvert\vect{h}_{kl}^{\Htran}\vect{D}_{il}\vect{w}_{il} \right\rvert^{2} \right\} -
	\begin{cases}
	0  & \text{for} \; i\neq k, \\
    \left\lvert a_{kl}\right\rvert^{2} & \text{for} \; i=k.
	\end{cases}
\end{align}
The max-min fairness optimization problem is then expressed in epi-graph form as
\begin{align}\label{eq:optimization formulation}
	\mathop{\mathrm{maximize}}\limits_{\{\gamma_{kl}:\forall k,l\}, s} \,\, & \quad s \\
	\text{subject to  }   & \, \mathrm{SINR}_{k}^{\mathrm{dl}}\geq s, \quad k=1,\ldots,K,\nonumber \\
	& \sum\limits_{i=1}^{K}\gamma_{il}^{2}\leq P^{\mathrm{dl}}_{\max}, \quad  l=1,\ldots,L .\nonumber 
\end{align}
The first constraint in \eqref{eq:optimization formulation} can be rewritten as
\begin{equation}\label{eq: SINR>S}
	\frac{1}{s}\left\lvert\sum\limits_{l=1}^{L}\gamma_{kl} a_{kl} \right\rvert^{2} \geq \sum\limits_{l=1}^{L}\sum\limits_{i=1}^{K}\gamma_{il}^{2} b_{kil} + \sigma^{2}.
\end{equation}
After taking the square root on both sides and noting that one can always rotate the phase of precoders to make $a_{lk} \geq 0$, (\ref{eq: SINR>S}) is rewritten as
\begin{equation}\label{eq:sqrt of SINR>S}
	\sqrt{\frac{1}{s}} \sum\limits_{l=1}^{L}\gamma_{kl} |a_{kl}| \geq \sqrt{\sum\limits_{l=1}^{L}\sum\limits_{i=1}^{K}\gamma_{il}^{2} b_{kil} + \sigma^{2}}
\end{equation}
and, equivalently, in vector form as
\begin{equation}\label{eq:SOCP form}
	\sqrt{\frac{1}{s}}\vect{c}^{\Ttran}_{k}\vect{\gamma}_{k} \geq \norm{\vect{B}_{k} \left[\begin{array}{c}
		\vect{\gamma}\\
		{\sigma}\\
		\end{array}\right]}
\end{equation}
where $\vect{c}_{k} = \left[
|a_{k1}| \, \ldots \, |a_{kL}| \right]^{\Ttran} \dimn{L\times 1}$, and $\vect{\gamma}_{k} = \left[
\gamma_{k1} \, \ldots \, \gamma_{kL} \right]^{\Ttran} \dimn{L\times 1}$. $\vect{B}_{k}=\diag\left(
\sqrt{b_{k11}}  \, \ldots \,\sqrt{b_{kKL}} \, \, 1 \right)\dimn{(KL+1)\times (KL+1)}$ and $\vect{\gamma} = \left[
\vect{\gamma}_{1}^{\Ttran} \, \ldots \, \vect{\gamma}_{K}^{\Ttran} \right]^{\Ttran} \dimn{KL\times 1}$. 

Hence, \eqref{eq:optimization formulation} can be equivalently written as
\begin{align}\label{eq:optimization formulation2}
	\mathop{\mathrm{maximize}}\limits_{\{\gamma_{kl}:\forall k,l\}, s} \quad s \\
	\text{subject to  } \quad\,\,\,   & 	\sqrt{\frac{1}{s}}\vect{c}^{\Ttran}_{k}\vect{\gamma}_{k} \geq \norm{\vect{B}_{k} \left[\begin{array}{c}
		\vect{\gamma}\\
		\sqrt{\sigma^{2}}\\
		\end{array}\right]}, \,\,\,\, k=1,\ldots,K,\nonumber \\
	& \sum\limits_{i=1}^{K}\gamma_{il}^{2}\leq P^{\mathrm{dl}}_{\max}, \quad  l=1,\ldots,L .\nonumber 
\end{align}
This is still  a non-convex problem but we notice that if $s$ is constant, the SINR constraint in (\ref{eq:SOCP form}) becomes a second-order cone (SOC) constraint. 
Hence, for a given $s$, the problem in \eqref{eq:optimization formulation2} becomes a second-order cone program (SOCP), which can be solved through the bisection method \cite{bjornson2013optimal} by considering a sequence of $s$ that converges to the global optimum. In each subproblem, the following problem must be solved:
\begin{align}\label{eq:optimization formulation3}
	\mathop{\mathrm{maximize}}\limits_{\{\gamma_{kl}:\forall k,l\},c} \quad c \\
	\text{subject to  } \quad\,\,\,   & 	\sqrt{\frac{1}{S}}\vect{c}^{\Ttran}_{k}\vect{\gamma}_{k} \geq \norm{\vect{B}_{k} \left[\begin{array}{c}
		\vect{\gamma}\\
		\sqrt{\sigma^{2}}\\
		\end{array}\right]}, \,\,\,\, k=1,\ldots,K\nonumber \\
	& \sum\limits_{i=1}^{K}\gamma_{il}^{2}\leq c P^{\mathrm{dl}}_{\max}, \quad  l=1,\ldots,L \nonumber 
\end{align}
and verified to be feasible and have $c\leq 1$ at the solution. Note that the power scaling variable $c$ is introduced to improve the algorithm convergence, as suggested in \cite{bjornson2013optimal}. If not included, the bisection algorithm requires an extremely high number of iterations (accuracy) to find the max-min solution.

\section{Neural Network based Power Allocation}
The central goal of this paper is to demonstrate that large-scale fading information is sufficient for computing the optimal powers. This is in contrast to the traditional optimization approach for solving \eqref{eq:optimization formulation3} that requires knowledge of $\{a_{kl}\}$ and $\{b_{kil}\}$. We advocate using the UEs' large-scale fading coefficients $\{\beta_{kl}\}$ to perform power allocation because they already capture the main feature of propagation channels and interference in the network, and can be easily measured in practice based on the received signal strength. Therefore, for a given location of APs, the problem is to learn the \emph{unknown} mapping between $\{\beta_{kl}\}$ 
and the optimal square-roots of the transmit powers $\{\gamma_{kl}^*\}$.
This is achieved by leveraging that DNNs are universal function approximators \cite{goodfellow2016deep,sun2018learning}.

We use a fully connected feed-forward DNN with $N$ hidden layers. The output layer provides an estimate $\{\hat{\gamma}_{kl}\}$ of $\{\gamma_{kl}^*\}$.
The problem is thus to train effectively the weights and biases of the DNN so that it can learn $\{\gamma_{kl}^*\}$. We consider two different DNNs with both MR and RZF precoding. The first is the so-called \emph{centralized} DNN that receives as input the entire large-scale fading coefficients $\{\beta_{kl}: \forall k,l\}$ and provides as output the network-wide square-roots of the powers $\{{\hat{\gamma}}_{kl}: \forall k,l\}$. 
The second DNN is called \emph{decentralized} because it operates on a per-AP basis. Specifically, the DNN of AP $l$ receives as input only the locally available coefficients $\{\beta_{kl}: \forall k\}$ and aims to learn the local estimate $\{{\hat{\gamma}}_{kl}: \forall k\}$ of optimal powers. The advantage of the decentralized DNN is that no exchange of large-scale fading coefficients among APs is required, which is important for a scalable network operation \cite{bjornson2019scalable}. Moreover, the number of trainable parameters per AP largely reduces. We stress that each AP has a unique DNN that captures features of the local propagation environment and where the AP is located compared to other APs.

\begin{figure}[!t]
	\centering\includegraphics[trim={5mm 2mm 5mm 5mm},width=\columnwidth]{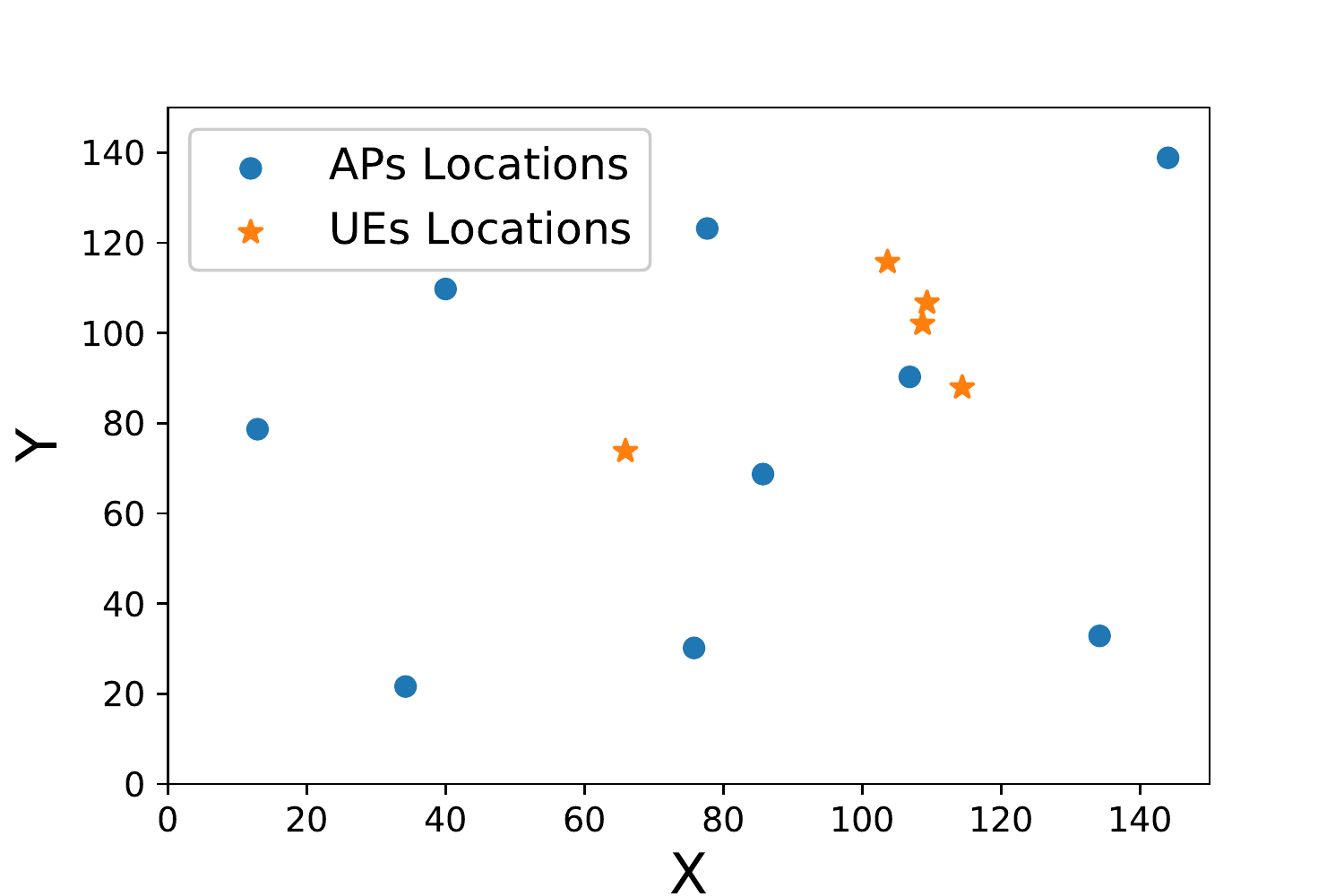}
	\caption{In the numerical evaluation, $L=9$ APs with fixed locations are considered as shown in this figure. In every user drop, $K=5$ UEs are dropped randomly in the area.}
	\label{fig:Random AP and UE distribution} \vspace{-4mm}
\end{figure}

The complexity of the approach above is mainly the generation of the training dataset. Suppose each layer of a DNN has $N_{i}$ neurons. The number of multiplications and addition for each layer is $N_{i}N_{i-1}$ and $2N_{i}, i=1,\ldots,N$, respectively. Each layer needs to evaluate $N_{i}$ activation functions. Once trained, the DNN allocates the transmit power very rapidly without actually solving \eqref{eq:optimization formulation3}. In practice, such decisions need to be made when the large-scale fading coefficients change due to large-scale movements, the addition of new UEs, or when UEs go from active to inactive mode.

\section{Numerical Evaluation}
\begin{table}[!t]
	\caption{Simulation parameters of the Cell-free mMIMO network}
	\begin{center} \vspace{-2mm}
		\label{table: Simulation parameters}
		\begin{tabular}{|c|c|}
			\hline
			Cell area (wrap around) & $150$ m$^{2}$ \\
			Bandwidth               &       $20$ MHz\\
			Number of APs           &       $L =9$\\
			Number of UEs           &      $K= 5$\\
			Number of antennas per AP & $M=2$\\
			Pathloss exponent ($\alpha$) & $\alpha =3.76$ \\
			Maximum downlink transmit power per AP & $P^{\mathrm{dl}}_{\max}=1$ W\\
			UL noise power & $-94$ dBm \\
			UL transmit power & $p_i=100$ mW \\
			Length of coherence block & $\tau_{c}=200$ \\
			\hline
		\end{tabular}
	\end{center}
\end{table}

To demonstrate the ability to learn how to perform power allocation based on only large-scale fading coefficients, we consider a Cell-free mMIMO network with $L=9$ APs at fixed locations in a square of $150$\,m $\times 150$\,m, as illustrated in Fig.~\ref{fig:Random AP and UE distribution}.  The large-scale fading coeffcients are generated as \cite{bjornson2019making}
\begin{align}\label{pathloss}
\beta_{kl}  = -30.5 - 36.7 \, \log_{10} \left( \frac{d_{kl}}{1\,\textrm{m}} \right)\quad \text{dB}
\end{align}
where $d_{kl}$ is the distance of UE $k$ from AP $l$.
In each realization of the network,  $K=5$ UEs are independently and random uniformly distributed in the area. The APs are deployed 10m above the UEs. All other simulation parameters are reported in Table \ref{table: Simulation parameters}.  The length of the pilot sequences is $\tau_{p}=K$, so orthogonal pilots are allocated to the UEs. We consider the downlink with MR or RZF precoding. 

\begin{table}[t!]
	\caption{Layout of centralized DNN for whole network with $L = 9$ and $K = 5$. Number of trainable parameters: 241965.}
	\begin{center} \vspace{-2mm}
		\label{table: Model CDNN}
		\begin{tabular}{l|l|l|l}
			& Size & Parameters & Activation Function \\  \hline 
			Input           & $KL$   &           & -                   \\
			Layer 1 (Dense) & 128  &5888     & elu                \\
			Layer 2 (Dense) &512  & 66048    & elu                \\
			Layer 3 (Dense) & 256  & 131328     & sigmoid               \\
			Layer 4 (Dense) & 128 & 32896      & sigmoid              \\
			Layer 5 (Dense) & $KL$ & 5805    & relu               
		\end{tabular}
	\end{center}
\end{table}
\begin{table}[t!]
	\caption{Layout of decentralized DNN for an AP with $L = 9$ and $K = 5$. Number of trainable parameters:  3877.}
	\begin{center} \vspace{-2mm}
		\label{table: Model DDNN}
		\begin{tabular}{l|l|l|l}
			& Size & Parameters & Activation Function \\  \hline 
			Input           & $K$   &           & -                   \\
			Layer 1 (Dense) & 16  & 96    & elu                \\
			Layer 2 (Dense) & 64  & 1088    & elu                \\
			Layer 3 (Dense) & 32  & 2080     & sigmoid               \\
			Layer 4 (Dense) & 16 & 528      & sigmoid              \\
			Layer 5 (Dense) & $K$ & 85    & relu               
		\end{tabular}
	\end{center} \vspace{-2mm}
\end{table}

The centralized and decentralized DNNs used with both precoding schemes are reported in Tables \ref{table: Model CDNN} and \ref{table: Model DDNN}, whose trainable parameters
are $241965$ and $3877$, respectively. The DNNs were trained based on a dataset of $N_{s} = 249900$ samples of independent realizations of the user locations, corresponding to large-scale fading coefficients $\{{\beta}_{kl}(n) : n=1,\ldots,N_s\}$ and the corresponding max-min power allocation variables $\{{{{\mu}}}_{kl}^{\star}(n) : n=1,\ldots,N_s\}$ for any given pair $k$ and $l$. Particularly, $90\%$ percent of the samples was used for training and $10\%$ for validation. The remaining $100$ samples formed the test dataset, which is independent from the training dataset and is used to generate the simulation results presented in this section. We used the Adam optimizer and categorical cross-entropy as loss function. The number of epochs, batch size and learning rate are optimized by a trial-and-error method.

The performance of both DNNs is evaluated by computing the cumulative distribution function (CDF) of the downlink SE per user, where the randomness is due to the UE locations. Comparisons are made with the performance achieved by the max-min optimization algorithm developed herein and the heuristic power allocation recently proposed in \cite{interdonato2019scalability}, where each AP uses full power and allocates power to UE proportionally to the square root of the channel estimation variance.

\begin{figure}[!t]
	\centering\includegraphics[width=9cm]{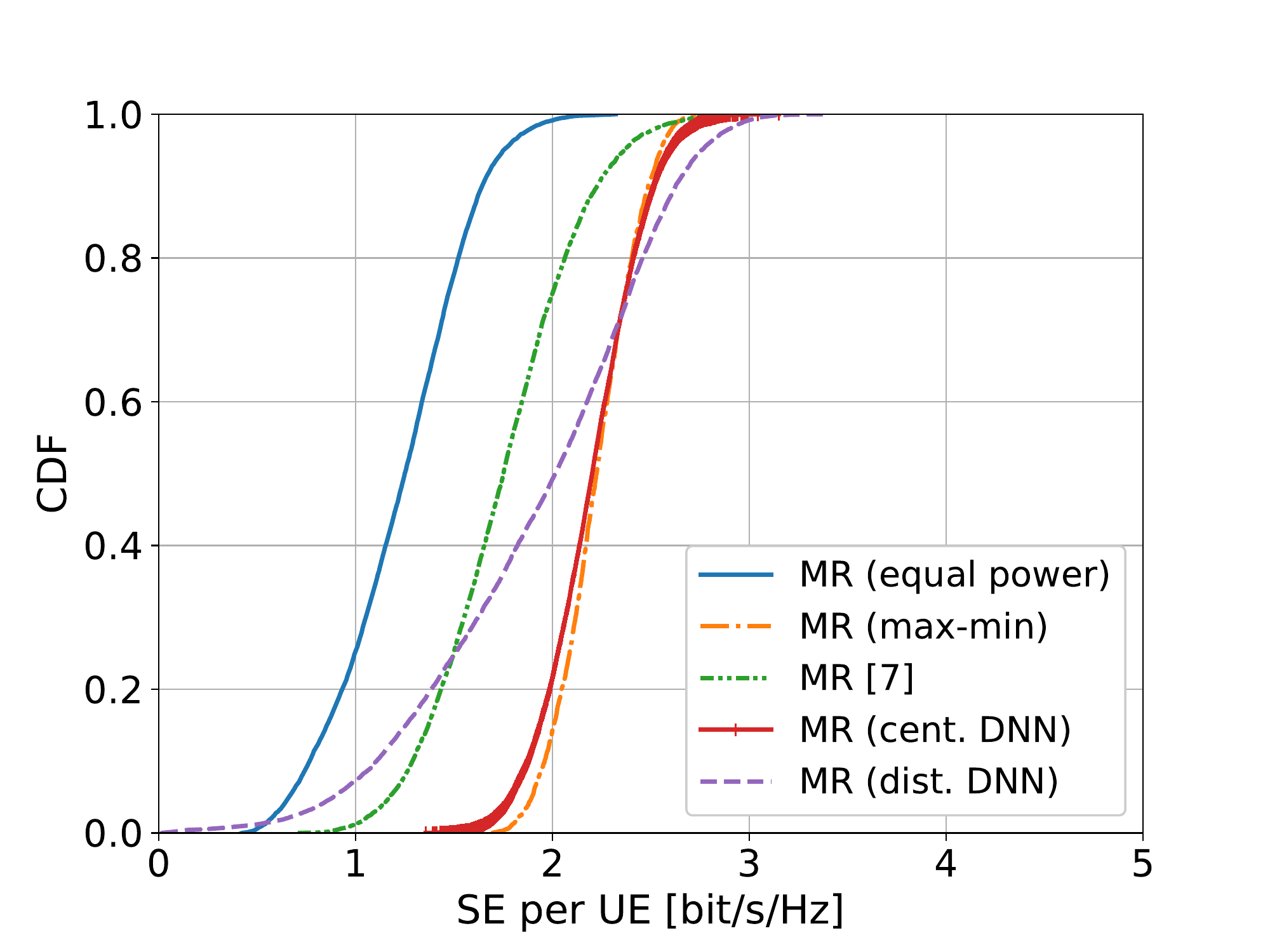}
	\caption{CDF of DL SE per UE with MR.}
	\label{fig:SE_CDF_maxmin_MR}
	\vspace{-0.2cm}
\end{figure}

Fig.~\ref{fig:SE_CDF_maxmin_MR} shows the CDF with MR precoding. The centralized DNN closely follows the performance achieved with the max-min algorithm.
The distributed DNN outperforms \cite{interdonato2019scalability} for 80\% of the UEs, but lower SE for the 20\% most unfortunate users. A possible explanation for the performance improvement using the DNN based power allocation is that it learns the  propagation environment of the network so that every AP can apply a locally optimized power allocation policy.
However, there is still a substantial gap between the distributed and centralized methods.

\begin{figure}[!t]
	\centering\includegraphics[width=9cm]{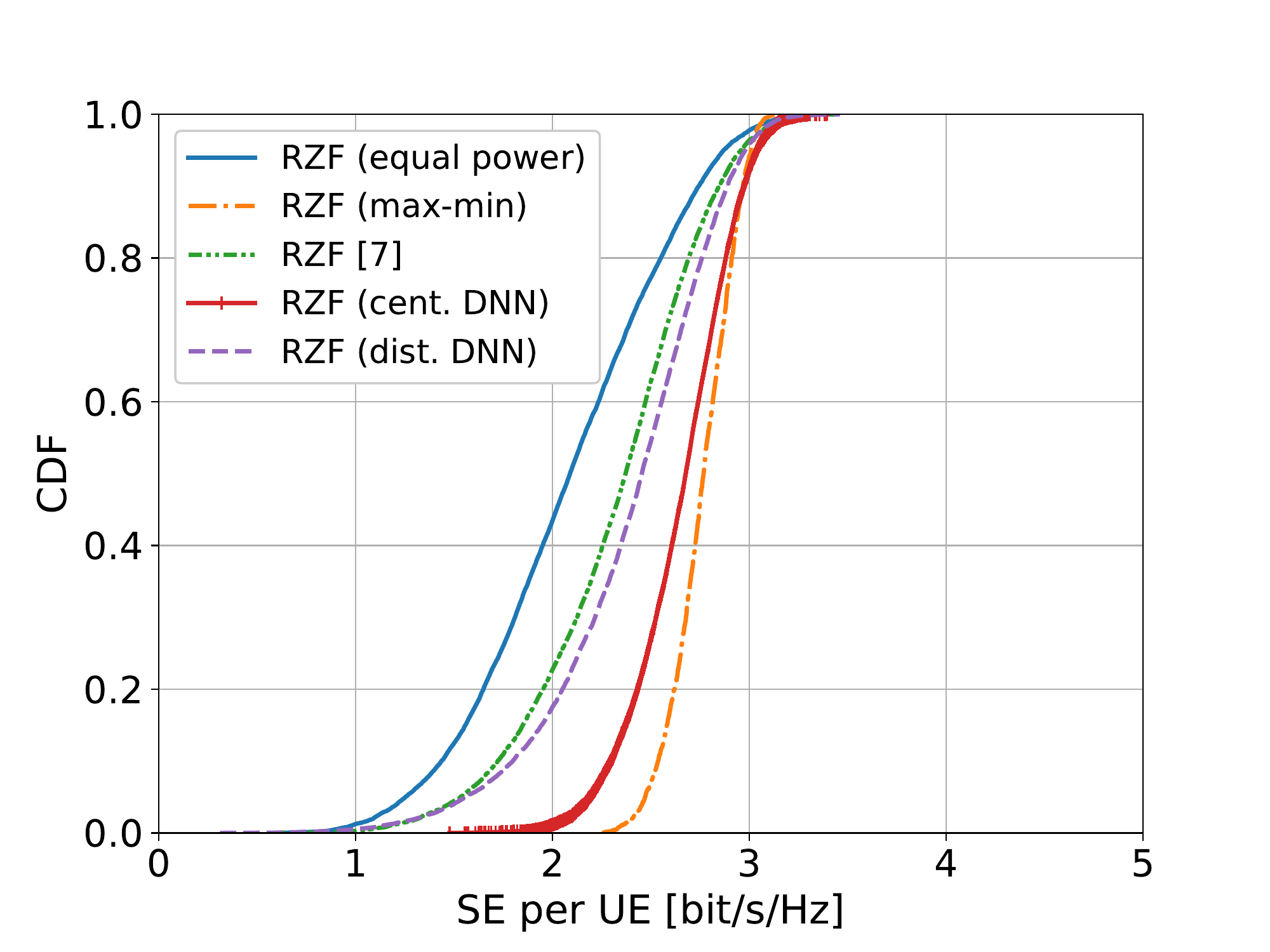}
	\caption{CDF of DL SE per UE with RZF.}
	\label{fig:SE_CDF_maxmin_RZF}
	\vspace{-0.2cm}
\end{figure}

Fig.~\ref{fig:SE_CDF_maxmin_RZF} shows the CDF of downlink SE per user with RZF precoding. For 40\% users, the centralized DNN power allocation closely approximate the conventional max-min fairness algorithm without having to actually solve (\ref{eq:optimization formulation3}). It achieves 42\% higher average SE than that of \cite{interdonato2019scalability}. In addition, local DNN at each AP achieves 4\% higher SE compared to \cite{interdonato2019scalability}. 

\section{Conclusions}
In this paper, we proposed a deep learning framework for downlink power allocation in a Cell-free mMIMO network with MR and RZF precoding and short-term power constraints. We developed the optimal power allocation strategy using the max-min fairness approach (which was previously only known for MR precoding with long-term power constraints) and used it to generate the training dataset for DNNs. We showed that a properly  trained feed-forward DNN is able to learn how to allocate powers. This is achieved by using only large-scale fading information, thereby substantially reducing the complexity and processing time of the optimization process. Also, we showed that a decentralized DNN at each AP can allocate power more effectively as compared to previous heuristic methods. However, there is still improvements to be made since the gap between the decentralized and centralized methods is rather large.

\bibliographystyle{IEEEtran}
\bibliography{IEEEabrv,references}

\end{document}